\documentstyle[psfig,aps,preprint]{revtex}

\def\beq{\begin{equation}}
\def\eeq{\end{equation}}
\def\bea{\begin{eqnarray}}
\def\eea{\end{eqnarray}}

\begin{document}

\draft

\title{Production of energetic dileptons with small invariant masses from the 
quark-gluon plasma\footnote{Supported by BMBF, GSI Darmstadt, and DFG}}
\author{Markus H. Thoma\footnote{Heisenberg fellow} and Christoph T. Traxler}
\address{Institut f\"ur Theoretische Physik, Universit\"at Giessen,\\
35392 Giessen, Germany}
\maketitle

\begin {abstract}

The resummation technique of Braaten and Pisarski is used for a complete calculation of the
$\alpha ^2\alpha _s$-contribution to the production rate of dileptons with energy
$E\gg T$ and invariant mass $M{\buildrel <\over \sim}T$ from a quark-gluon plasma.
In particular the cancellation of the infrared singularity by medium effects is discussed
in detail. 

\end{abstract}



\section{Introduction}

The thermal radiation of photons and dileptons from a quark-gluon plasma (QGP) is one of 
the most promising signatures for the formation of a QGP in relativistic heavy ion collisions
\cite{Ruus}. Whereas real photons can be produced to lowest order $\alpha \alpha _s$ only via
the participation of a gluon, dileptons are produced to lowest order $\alpha ^2$ by 
quark-antiquark annihilation into a virtual photon (Born term) \cite{Born}. However, in the
case of a small invariant photon mass $M^2=E^2-p^2$, where $E$ is the energy and $p=|{\bf p}|$
the momentum of the photon, radiative corrections of the order $\alpha ^2 \alpha _s$ become
increasingly important. These corrections have been considered 
using perturbative QCD at finite temperature \cite{Corr,Alru}. In contrast to the production
of real photons the dilepton production rate turns out to be infrared finite in the case of a
vanishing quark mass. There is a cancellation of the infrared singularities of real and
virtual contributions, the latter appearing only for the production of virtual photons
\cite{Alru}. However, also medium effects leading to an effective quark mass $m_q^*$
by the interaction of the quark with the heat bath can screen these infrared divergences.
Altherr and Ruuskanen \cite{Alru} suggested that the effective quark mass simply replaces the 
invariant photon mass as infrared cutoff if $M<m_q^*$. 

Medium effects can be included consistently using the resummation technique of Braaten
and Pisarski \cite{Brpi}. This method has been applied to a number of interesting
quantities of the QGP leading to gauge invariant and infrared finite results that are complete
to leading order in the coupling constant \cite{Thom}. For example, the production rate
of energetic photons has been derived in this way, where a resummed quark propagator
containing the effective quark mass ${m_q^*}^2=g^2T^2/6$ has been used \cite{Kls,Bai,Tvt}.
Here we want to reconsider the result of Altherr and Ruuskanen \cite{Alru} performing
a complete calculation of the dilepton rate using the Braaten-Pisarski method analogously
to the photon case. In particular we will study in detail the role of $m_q^*$ and $M$ as 
infrared cutoffs. Like Altherr and Ruuskanen, we will restrict ourselves
to small photon masses $M{\buildrel <\over \sim}T$, for which the $\alpha _s$-corrections 
are of importance, and to large photon energies $E\gg T$. 

\section{Dilepton production rate to order {\boldmath $\alpha ^2 \alpha 
\mbox{\unboldmath $_s$}$ \unboldmath}}

The lowest order matrix elements for the production of virtual photons from the QGP are 
shown in Fig.1. Besides the real contributions (Compton scattering and annihilation) of
Fig.1a,b, appearing also for the production of real photons, there are virtual
contributions, namely gluon absorption (Fig.1c) and the interference terms of the Born
term with radiative corrections (self energy insertion and vertex correction) in Fig.1d.
Due to phase space restrictions and kinematics ($E\gg T$) the gluon absorption
as well as the vertex correction can be neglected \cite{Alru}. 

The matrix elements can be related to the imaginary part of the photon self energy
diagrams in Fig.2 \cite{Weld}. Here the processes of Fig.1a,b,c can be obtained by cutting
through the internal quark and gluon lines, while the interference term (Fig.1d)
corresponds to a cut through the quark lines only. In the latter case the quark 
self energy becomes on-shell, leading to an infrared singularity which cancels the one
from the exchanged massless quark in the  real contribution \cite{Alru}. This cancellation
can be considered as an example of the KLN-theorem at finite temperature \cite{Kln}.

The production rate for massless electron and muon pairs derived from the photon self energy is 
given by \cite{Alru}
\beq
\frac {dR}{d^4xd^4p}=\frac {1}{6\pi ^4}\> \frac {\alpha }{M^2}\> \frac {1}{e^{E/T}-1}\>
Im \Pi _\mu ^\mu (P).
\label{e1}
\eeq

As in the case of the photon production rate \cite{Kls,Bai,Tvt}, we want to include 
the effect of the medium by using a resummed quark propagator for soft momenta of the 
exchanged quark. For this purpose we introduce a separation scale $k_c$ \cite{Bryu},
restricted by $gT\ll k_c\ll T$ in the weak coupling limit. For quark momenta smaller than 
$k_c$ we will start from the photon self energy shown in Fig.3, where the blob denotes
the effective quark propagator, in which the quark self energy in the hard thermal loop
approximation has been resummed \cite{Bpy}. Owing to the high energy of the photon and
energy-momentum conservation we need to take into account only one effective quark 
propagator and no effective vertices as opposed to the case of soft dileptons \cite{Bpy}.

For momenta of the exchanged quark larger than $k_c$, the diagrams of Fig.1 or Fig.2,
containing only bare propagators, will be considered. After adding up the
soft and hard contributions, the separation scale has to drop out, demonstrating
the completeness of the calculation \cite{Bryu}. In the case of the photon production
a covariant separation scale, i.e. $|K^2|=|\omega ^2-k^2|=k_c^2$ \cite{Kls}, and a
non-covariant one ($k=k_c$) \cite{Bai} have been used, both leading to the same result.

In the limit $M^2\ll ET$ the hard real contribution (Fig.1a,b) of the dilepton production rate
can be taken over from the photon rate \cite{Alru}, yielding in the case of a non-covariant
cutoff \cite{Bai}
\beq
{\left (\frac {dR}{d^4xd^4p}\right )}_{hard}^{real}
=\frac {10}{27\pi ^3}\> \alpha ^2\alpha _s \> \frac
{T^2}{M^2}\> e^{-E/T}\> \left [\ln \frac {ET}{k_c^2}+\frac{3}{2}+\frac {\ln 2}{3}-\gamma
+\frac {\zeta '(2)}{\zeta (2)}\right ],
\label{e2}
\eeq
where $\gamma =0.57722$ and $\zeta '(2)/\zeta (2)=-0.56996$. In the case of a covariant cutoff
the constant in the square brackets is changed by adding a term $\ln 4-2$ \cite{Kls,Bai}.
The hard contribution to the photon rate follows from a momentum integration over the square 
of the matrix elements of Fig.1a,b using distribution functions for the external partons.
Applying the Boltzmann approximation for the distributions of the incoming partons
and using $k_c\ll T$ the result (\ref{e2}) is found after some tedious manipulations.
It can be obtained more easily by calculating the photon absorption rate and applying
the principle of detailed balance \cite{tho1}. 

The soft contribution to the dilepton rate can be computed by introducing spectral functions
for the resummed quark propagator \cite{Bpy} as in the case of the photon rate \cite{Kls}.
Following the arguments in the calculation of the photon rate we arrive at
\bea
{\left (\frac {dR}{d^4xd^4p}\right )}_{soft}^{real} & = & \frac {40}{9\pi ^2}\> \frac {\alpha
^2}{M^2}\> \int \frac {d^3k}{(2\pi )^3}\> \int d\omega \> \int d\omega '\> \delta (E-\omega
-\omega ')\> n_F(\omega )\> n_F (\omega ')\nonumber \\
&& \biggl \{ \left (1+\hat {\bf q}\cdot \hat {\bf k}\right )\> \left [\rho _+(\omega ,k)\,
\delta (\omega '+q)+\rho _-(\omega ,k)\, \delta (\omega '-q)\right ]\nonumber \\
&&{}+
\left (1-\hat {\bf q}\cdot \hat {\bf k}\right )\> \left [\rho _+(\omega ,k)\,
\delta (\omega '-q)+\rho _-(\omega ,k)\, \delta (\omega '+q)\right ]\biggr \},
\label{e3}
\eea
where $n_F$ is the Fermi-Dirac distribution, ${\bf q}={\bf p}-{\bf k}$ the momentum of the
bare quark propagator, and
\beq
\rho _\pm (\omega, k)=\frac {\omega ^2-k^2}{2{m_q^*}^2}\> [\delta (\omega -\omega _\pm )+
\delta (\omega +\omega _\pm )]+\beta _\pm (\omega , k)\> \theta (k^2-\omega ^2)
\label{e4}
\eeq
the spectral functions with
\beq
\beta _\pm (\omega ,k)=-\frac {{m_q^*}^2}{2}\> \frac {\pm \omega -k}{\left [k\, (-\omega \pm k)
+{m_q^*}^2\, \left (\pm 1-\frac {\pm \omega -k}{2k}\, \ln \frac {k+\omega }{k-\omega } \right )
\right ]^2+\left [\frac {\pi}{2}\, {m_q^*}^2\, \frac {\pm \omega -k}{k}\right ]^2}.
\label{e5}
\eeq
The first term of the spectral functions (\ref{e4}), containing the dispersion relations 
$\omega _\pm  (k)$ (see Fig.4) of the collective quark modes \cite{Klwe} above the light
cone ($\omega >k$), gives rise to the virtual contribution of the dilepton rate. The real soft
contribution, on the other hand, follows from the second term below the light cone
($-k<\omega <k$), which comes from the imaginary part of the hard thermal loop quark self
energy. Using $E\gg T$ and assuming $\omega $ and $k$ to be soft \cite{Kls,Tvt}, the soft
real contribution reduces to
\bea
{\left (\frac {dR}{d^4xd^4p}\right )}_{soft}^{real}
=\frac {5}{9\pi ^4}\> \frac {\alpha ^2}{M^2}\> e^{-E/T}\>
\int dk d\omega \> &&{} [(k-\omega +E-p)\> \beta _+(\omega ,k)\nonumber \\
&&{} +(k+\omega -E+p)\> \beta _-(\omega ,k)].
\label{e6}
\eea
The integration range, determined by $\omega =-k+E-p$, $\omega
=k$, and the separation scale $k=k_c$ or $k^2-\omega ^2=k_c^2$, respectively, are
shown in Fig.4. Compared to the photon case it is shifted by $E-p>0$, which renders
the integration in (\ref{e6}) more difficult. 
The logarithmic dependence on the scale $k_c$ can be extracted by adopting the static limit
of the effective quark propagator \cite{Bai},
\beq
\beta _\pm (\omega =0,k)=\frac {{m_q^*}^2}{2}\> \frac{k}{(k^2+{m_q^*}^2)^2+(\pi {m_q^*}^2/2)^2}.
\label{e7}
\eeq
In the case of a non-covariant separation scale $k_c$, where the $\omega $-integration 
is restricted by the limits $-k+E-p$ and $k$, while the $k$-integration ranges from 
$(E-p)/2$ to $k_c$, we find to logarithmic approximation
\beq
{\left (\frac {dR}{d^4xd^4p}\right )}_{soft}^{real}
=\frac {5}{9\pi ^4}\> \frac {\alpha ^2}{M^2}\> e^{-E/T}\> {m_q^*}^2\> \left (\ln 
\frac {k_c^2}{{m_q^*}^2}+A\right ),
\label{e8}
\eeq
where we have used $k_c\gg gT$. In order to determine the function $A(E-p)$ beyond the 
logarithm, 
we cannot use the static approximation anymore, but have to solve (\ref{e6}) together with 
(\ref{e5}) numerically. We restrict ourselves to $E-p\leq m_q^*$, which holds for realistic
values of the coupling constant $g{\buildrel > \over \sim }1$. Then we find that $A$ increases
from $A(E-p=0)=-1.31$, which corresponds to the photon result \cite{Bai}, to 
$A(E-p=m_q^*)=-1.71$.

Adding the hard contribution (\ref{e2}) to (\ref{e8}) with ${m_q^*}^2=2\pi \alpha_s T^2/3$, 
the arbitrary separation scale $k_c$ drops out
and we obtain an infrared finite result for the real contribution of the dilepton rate.
On the other hand, using a covariant separation scale, (\ref{e8}) is multiplied by a factor
$k_c^2/[(E-p)^2+k_c^2]$. Note that $E-p=M^2/(E+p)\ll T$ might be of the 
same order as $k_c\ll T$ for $M\sim T$. 
Hence the separation scale does not cancel in this case in contrast
to the real photon rate ($E=p$). This shows that we should demand that $\omega $
and $k$ are soft individually as it was also assumed in the derivation of the
Braaten-Pisarski method, which is based on the imaginary time formalism in euclidean space-time
\cite{Brpi}.

Finally, we consider the virtual part coming from the pole contribution to the imaginary 
part of the photon self energy in (\ref{e1}). Now we have to use a covariant separation scale 
since the infrared  singularity in Fig.2 comes from the on-shell self energy at $K^2=0$. 
The soft part follows from the pole contribution of the diagrams in Fig.3
corresponding to the first term of the spectral functions (\ref{e4}). Since 
$\omega _\pm (k)$ lies always below $K^2=k_c^2\gg {m_q^*}^2$ (see Fig.4), 
there is no hard contribution 
to the virtual part. Thus the virtual contribution is determined solely by the imaginary 
part of the self energy in Fig.3 coming from the pole of the effective quark propagator,
which gives a finite result when integrated over the entire momentum range. 

Combining (\ref{e3}) and the first term of (\ref{e4}) we find
\bea
{\left (\frac {dR}{d^4xd^4p}\right )}^{virt} 
& = & \frac {5}{9\pi ^4}\> \frac {\alpha ^2 }{M^2 {m_q^*}^2}\> \frac {1}{p}\>
\nonumber \\
\biggl \{ \! \! \! \! \! \! \! \! &&{} \int _{k_{min}^+}^{k_{max}^+} dk\>  
n_F(\omega _+)\> n_F(E-\omega _+)\> (\omega _+^2-k^2)\> \left [ \frac {M^2}{2}-E\, 
(\omega _+-k)+\frac {(\omega _+-k)^2}{2}\right ]\nonumber \\
& - & \int _{k_{min}^-}^{k_{max}^-} dk\>  n_F(\omega _-)\> n_F(E-\omega _-)\>
(\omega _-^2-k^2)\> \left [ \frac {M^2}{2}-E\, (\omega _-+k)
+\frac {(\omega _--k)^2}{2}\right ] \biggr \},\nonumber \\
\label{e9}
\eea
where the limits $k_{max,min}^\pm$ are determined from the intersections of the 
dispersion relations with $\omega =\pm k\pm (E-p)$ as shown in Fig.4. Note that 
the virtual part vanishes for $M\rightarrow 0$ as $k_{min}^\pm $ tends to infinity. 

In order to isolate the term proportional to $\alpha ^2\alpha _s$ and to investigate its 
dependence on $M$, $m_q^*$, $E$, and $T$, we introduce again a separation scale 
$gT\ll k_s\ll T$. For $k<k_s$ we may approximate the distribution functions by
$n_F(\omega _\pm)\simeq 1/2$ and $n_F(E-\omega _\pm)\simeq \exp(-E/T)$, whereas
for $k>k_s$ we may set $\omega _+\simeq k+{m_q^*}^2/k$ leading to $\omega _+^2-k^2
\simeq 2{m_q^*}^2$ and $\omega _-=k$ \cite{Pis}. Hence the plasmino branch $\omega _-(k)$
does not contribute for hard momenta as $\omega _-$ approaches $k$ exponentially
for $k\gg gT$ \cite{Pis}.  

For the part of the integrals in (\ref{e9}) containing the terms with $M^2/2$ it is
sufficient to restrict to $k>k_s$ since the hard part is finite for $k_s \rightarrow 0$. 
Considering $E\gg T$ the simplifications 
$k_{max}^+\simeq (E+p)/2$, $k_{min}^+\simeq (E-p)/2$, $n_F(\omega _+)\simeq n_F(k)$, and
$n_F(E-\omega _+)\simeq n_F(E-k)$ can be assumed. Then this term reduces to the Born term,
which reads for $E\gg T$ and $E-p\ll T$
\beq
{\left (\frac {dR}{d^4xd^4p}\right )}^{Born}=\frac {5}{9\pi ^4}\> \alpha ^2 \> e^{-E/T}.
\label{e10}
\eeq
The Born term is contained in the virtual contribution (\ref{e9}) because the effective
quark propagator includes the bare one.

Next we consider the terms under the integral in (\ref{e9}) proportional to $E(\omega _\pm 
\pm k)$. Using the approximations discussed above, we find for $k<k_s$
\bea
{\left (\frac {dR}{d^4xd^4p}\right )}^{virt}_{soft} & = & \frac {5}{18\pi ^4}\> \frac 
{\alpha ^2 }{M^2 {m_q^*}^2}\> \frac {1}{p}\> e^{-E/T}\> \nonumber \\
&& \left [-\int _{k_{min}^+}^{k_s} dk\> (\omega _+^2-k^2)\> (\omega _+-k)
       +\int _{k_{min}^-}^{k_s} dk\> (\omega _-^2-k^2)\> (\omega _-+k)\right ].
\label{e11}
\eea
We proceed analogously to Baier et al. \cite{Bai} in the case of the photon rate, using
$(\omega _\pm^2-k^2)(\omega _\pm \mp k)/{m_q^*}^2=\omega _\pm -k(d\omega _\pm/dk)$. Then we obtain with $k_s\gg m_q^*$
\beq
{\left (\frac {dR}{d^4xd^4p}\right )}^{virt}_{soft}=-\frac {5}{9\pi ^4}\> \frac 
{\alpha ^2 }{M^2}\> e^{-E/T}\> {m_q^*}^2\> \left (\ln \frac {k_s}{m_q+k_{min}^+} +B\right ).
\label{e12}
\eeq
The function $B(E-p)$ decreases from $B(E-p=0)=0$, where the virtual contribution vanishes, 
to $B(E-p=m_q^*)=-0.66$, where $k_{min}^\pm=0$ (see Fig.4).

The hard part, $k>k_s$, using the approximations given above, reads
\beq
{\left (\frac {dR}{d^4xd^4p}\right )}^{virt}_{hard}=-\frac {10}{9\pi ^4}\> \frac 
{\alpha ^2 }{M^2}\>  {m_q^*}^2\> \int _{k_s}^E dk\> \frac {n_F(k)n_F(E-k)}{k}.
\label{e13}
\eeq
Considering $k_s\ll T\ll E$, we find
\beq
{\left (\frac {dR}{d^4xd^4p}\right )}^{virt}_{hard}=-\frac {5}{9\pi ^4}\> \frac 
{\alpha ^2 }{M^2}\> e^{-E/T}\> {m_q^*}^2\> \ln \frac {E}{k_s}.
\label{e14}
\eeq

Adding up both the contributions (\ref{e12}) and (\ref{e14}) $k_s$ cancels and we arrive at
\beq
{\left (\frac {dR}{d^4xd^4p}\right )}^{virt}=-\frac {5}{9\pi ^4}\> \frac {\alpha ^2 }{M^2}
\> e^{-E/T}\> {m_q^*}^2\> \left (\ln \frac {E}{m_q^*+k_{min}^+}+B\right ).
\label{e15}
\eeq
We observe that the virtual contribution coming from an interference term is negative,
thus reducing the dilepton rate compared to the photon rate. Furthermore, the virtual
contribution vanishes for $m_q^*\rightarrow 0$ as it is also the case for the virtual 
contribution in naive perturbation theory \cite{Alru}.

The part under the integral in (\ref{e9}) proportional to $(\omega _\pm -k)^2$ can be shown 
to be suppressed by $m_q^*/E$ relative to (\ref{e15}).
 
Combining (\ref{e2}), (\ref{e8}), and (\ref{e15}) we end up with our final result for
the dilepton production rate
\beq
\frac {dR}{d^4xd^4p}=\frac {10}{27\pi ^3}\> \alpha ^2\alpha _s \frac {T^2}{M^2}\> 
e^{-E/T}\> \left (\ln \frac {T(m_q^*+k_{min}^+)}{{m_q^*}^2}+C\right ),
\label{e16}
\eeq
where the function $C$ ranges from $C(E-p=0)=-0.73$ to $C(E-p=m_q^*)=-0.47$.

For giving an approximate analytic expression for $k_{min}^+$ we replace the exact
dispersion relation $\omega _+(k)$ by $\omega _+^2=k^2+{m_q^*}^2$, which deviates from 
the exact one by less than 11\% over the entire momentum range. Then we obtain
$k_{min}^+=|E{m_q^*}^2/M^2-M^2/(4E)|$, restricted by $k_{min}^+\leq k_{max}^+\simeq (E+p)/2$.

\section{Discussion}

Now we will discuss our result (\ref{e16}) for various limits of $M$. For
$M\rightarrow 0$ the virtual contribution vanishes since $k_{min}^+$ tends to infinity
for $E-p\rightarrow 0$. Hence the dilepton rate is given by the real contribution which
agrees with the photon rate in this limit.

For $M\sim m_q^*$ we get $E-p=M^2/(2E)\ll m_q^*$. Hence $k_{min}^+$ is given by the 
intersection
of $\omega \simeq k+{m_q^*}^2/k$ and $\omega=k+E-p$, i.e. $k_{min}^+={m_q^*}^2/(E-p)
\sim E$. Thus the result for the photon rate holds approximately, which agrees
with the result of Altherr and Ruuskanen \cite{Alru}, if we replace the invariant
photon mass in the logarithmic term of their formula (2.18) by the effective quark mass
in the case of $M<m_q^*$ as suggested in their paper. 

For $M\sim T$, i.e. $E-p\sim T^2/E$, we have to distinguish two cases. First $E-p\sim m_q^*$,
which leads to $k_{min}^+\sim m_q^*$ (see Fig.4). Then the logarithm in (\ref{e16}) is given by
$\ln (T/m_q^*)$. Secondly, in the case $E-p\ll m_q$ 
we find from Fig.4 $k_{min}^+\gg m_q^*$. Now if $k_{min}^+\sim E$ we recover the photon result, 
while for $k_{min}^+\sim T$ the logarithmic term is given by $\ln (T^2/{m_q^*}^2)$. 

Finally we will comment on the extrapolation of our result to realistic values of the coupling 
constant, $\alpha _s=0.2$-0.5, i.e. $g=1.5$-2.5. In any case the lowest order result
in $\alpha _s$ fails if the rate (\ref{e16}) becomes negative. (This unphysical
behaviour can originate from extracting the leading order contribution and extrapolating to 
large
values of the coupling constant. A similar problem has been encountered in the case of transport 
rates determining thermalization times and the viscosity of the QGP \cite{Thom,Tho2}.)
Thus one might expect 
that for $M\sim T$ the result gets unphysical, since $m_q^*=T$ for $g=\sqrt {6}$.
However, for $E\gg T$ we find $E-p\ll T\sim m_q^*$ and $k_{min}^+\gg T$. 
Hence we end up with a logarithmic term which is always positive. However, the constant
$C$ behind the logarithm is of the same order (typically half of the value of the logarithm),
indicating the necessity to go beyond the logarithmic approximation. (This is more important 
for dileptons than for photons due to the negative virtual contribution.). 

As an example we have chosen $\alpha _s=0.3$ and $T=300$ MeV. In Fig.5 the dilepton 
production rate is shown as a function of $E$ for $M=300$ MeV and in Fig.6 as a function 
of $M$ for $E=3$ GeV. Also shown is the result of Altherr and Ruuskanen \cite{Alru}.
(Here we replaced $M$ in their formula (2.18) by $m_q^*$ for $M<m_q^*$.)
For the chosen set of parameters both approaches give similar results, since for
$M<m_q^*=238$ MeV both rates are given by the photon result approximately (see Fig.6).  
However, the dilepton rate found by Altherr and Ruuskanen breaks down  at somewhat larger 
values of $E$ as shown in Fig.5. In Fig.5 also the Born term (\ref{e10}) is depicted for
comparision.

Summarizing, we conclude that the complete calculation using a resummed quark propagator
according to the Braaten-Pisarski method leads to a finite result for the dilepton rate
to order $\alpha ^2\alpha _s$, where the effective quark mass $m_q^*$ cuts off the infrared
divergence. There is no need for using the invariant photon mass $M$ as an infrared
cutoff even for $M$ of the order of the temperature. 

\acknowledgements

We would like to thank E. Braaten and V. Ruuskanen for helpful discussions.

\begin{figure}
\centerline{\psfig{figure=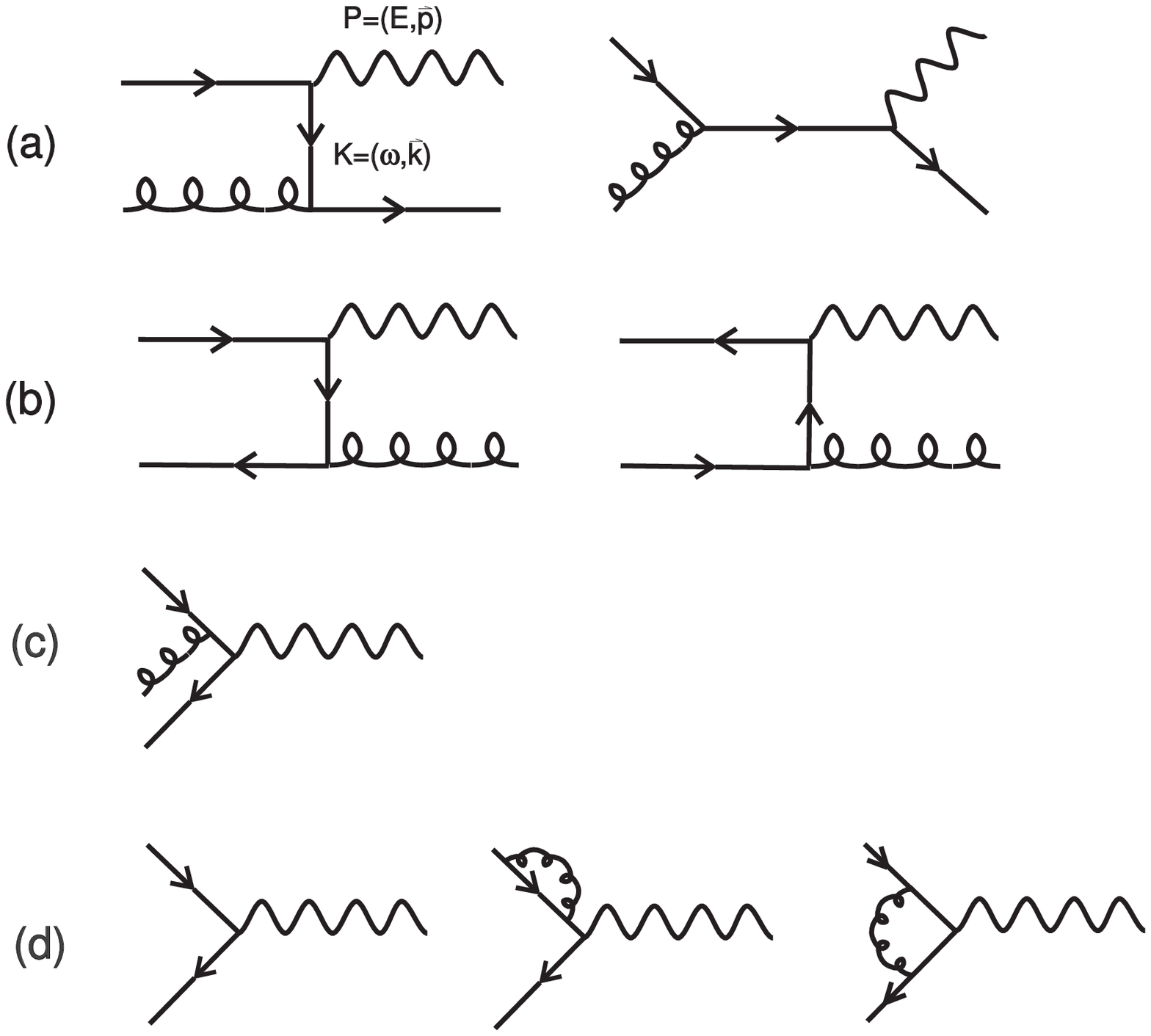,height=12cm}}
\caption{Matrix elements for dilepton production to order $\alpha ^2\alpha _s$.}
\end{figure}

\begin{figure}
\centerline{\psfig{figure=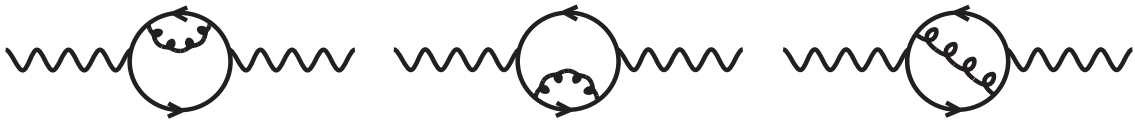,width=15cm}}
\caption{Photon self energy diagrams related to the matrix elements of Fig.1 by cutting rules.}
\end{figure}

\begin{figure}
\centerline{\psfig{figure=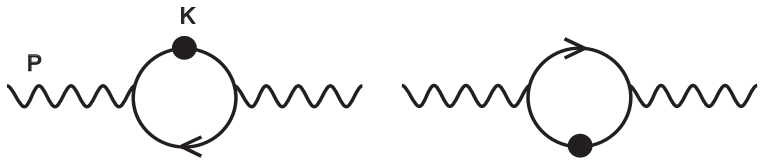,width=10cm}}
\vspace{0.5cm}
\caption{Photon self energy diagrams with an effective quark propagator.}
\end{figure}

\begin{figure}
\centerline{\psfig{figure=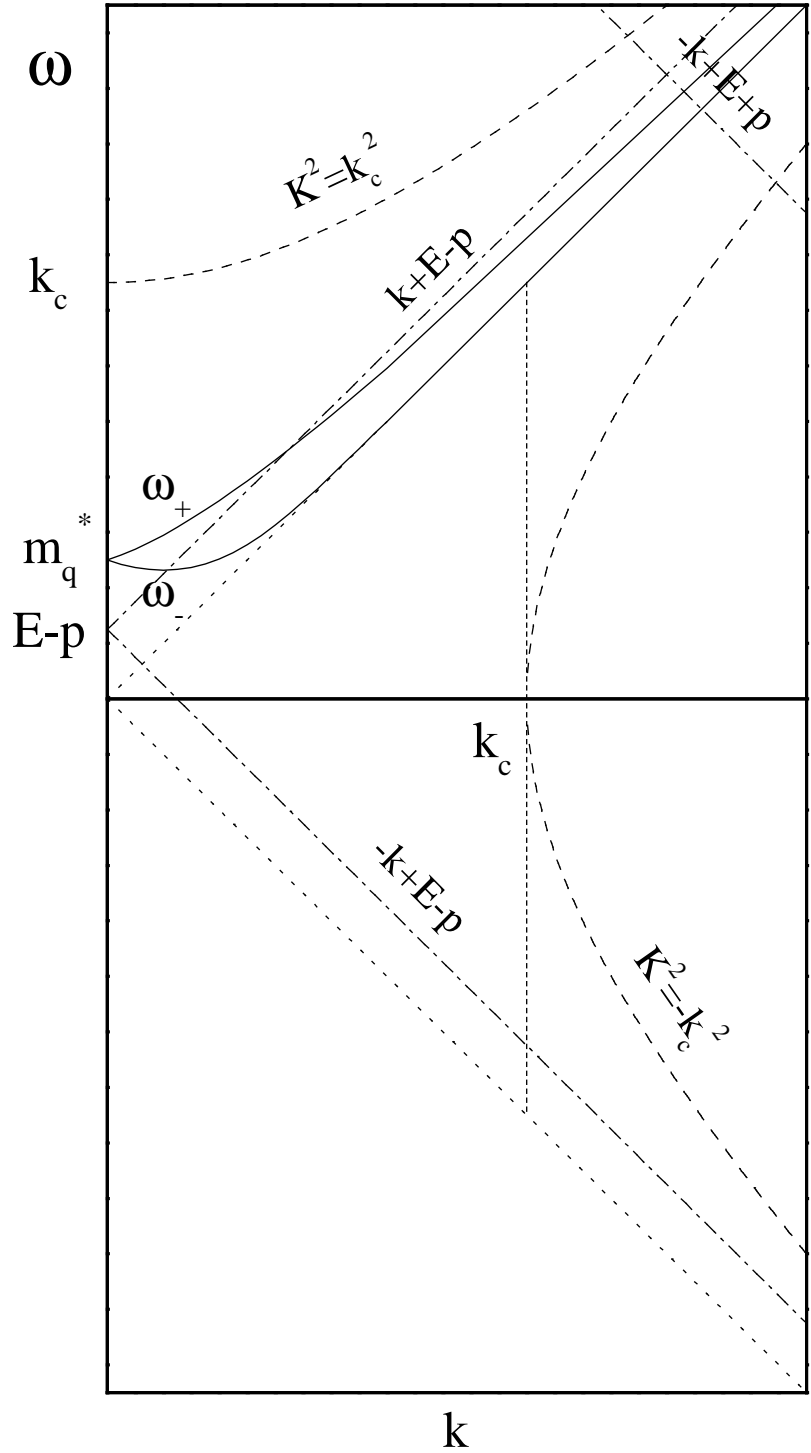,height=20cm}}
\caption{Energy $\omega $ and momentum $k$ of the exchanged quark restricted by kinematics
and the separation scale $k_c$. The real part of the dilepton rate arises from the region below
the light cone, whereas the virtual part follows from the quark dispersion relations 
$\omega _\pm (k)$.}
\end{figure}

\begin{figure}
\centerline{\psfig{figure=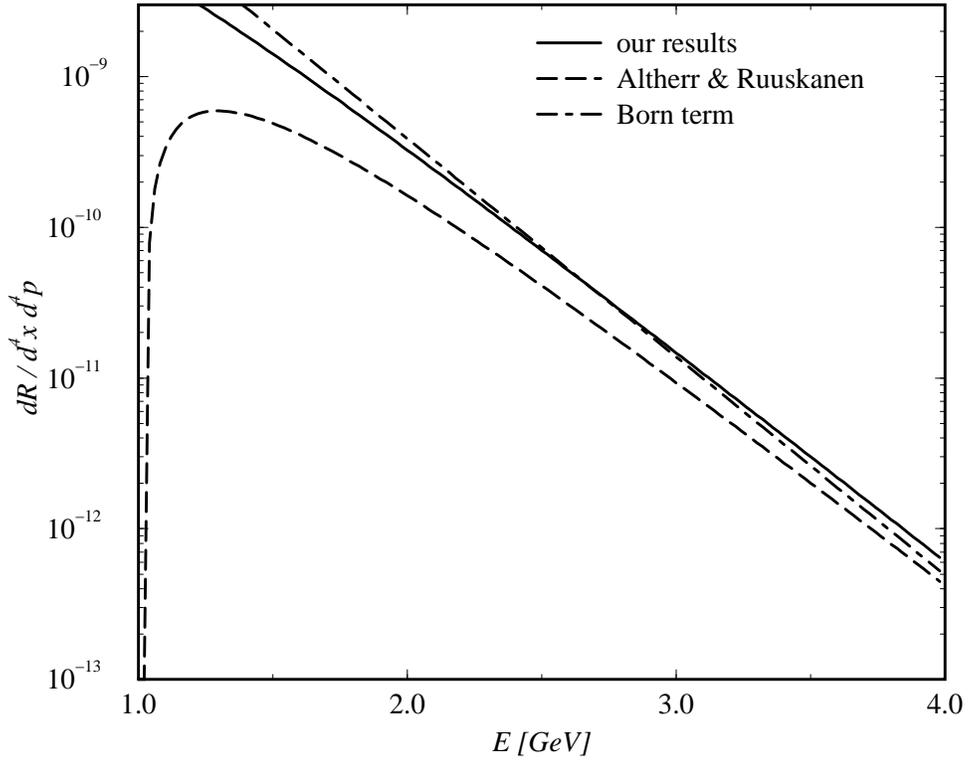,height=12cm}}
\caption{Contribution of order $\alpha ^2\alpha _s$ to the dilepton production rate
versus the energy of the virtual photon $E$ for $T=300$ MeV, $\alpha _s=0.3$, and $M=300$ MeV.}
\end{figure}

\begin{figure}
\centerline{\psfig{figure=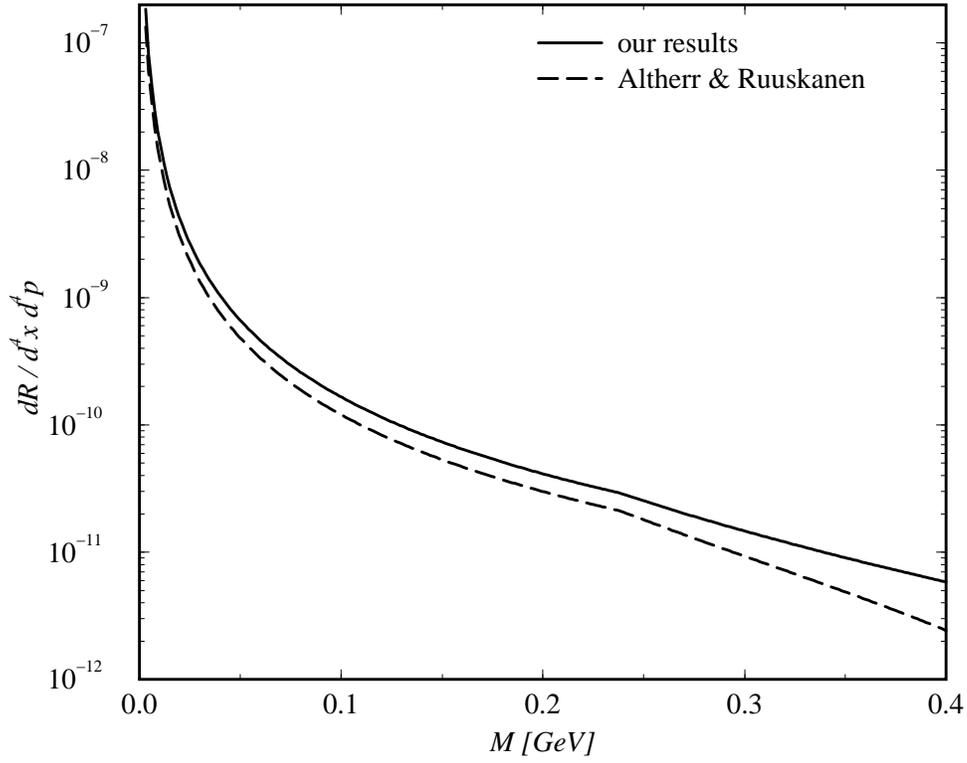,height=12cm}}
\caption{Contribution of order $\alpha ^2\alpha _s$ to the dilepton production rate
versus the invariant photon mass $M$ for $T=300$ MeV, $\alpha _s=0.3$, and $E=3$ GeV.}
\end{figure}

\end{document}